\documentclass[aps,pra,reprint,showpacs,groupedaddress, twocolumn]{revtex4-1}
\usepackage{amssymb}
\usepackage{mathrsfs}
\usepackage{amsmath}
\usepackage{amsfonts}
\usepackage{graphicx}
\usepackage{dcolumn}
\usepackage{color}
\usepackage{subfigure}
\usepackage{epsfig}
\usepackage{soul}
\usepackage[colorlinks,citecolor=blue, linkcolor=blue,hyperindex,CJKbookmarks,dvipdfm]{hyperref}

\begin{document}

\title{Optimal unidirectional amplification induced by optical gain \\
in optomechanical systems}
\date{\today }
\author{L. N. Song$^{1}$}
\author{Qiang Zheng$^{2,1}$}
\email{qz@gznu.edu.cn}
\author{Xun-Wei Xu$^{3}$}
\author{Cheng Jiang$^{1,4}$}
\author{Yong Li$^{1,5}$}
\email{liyong@csrc.ac.cn}

\begin{abstract}
We propose a three-mode optomechanical system to realize optical
nonreciprocal transmission with unidirectional amplification, where the
system consists of two coupled cavities and one mechanical resonator which
interacts with only one of the cavities. Additionally, the optical gain is
introduced into the optomechanical cavity. It is found that for a strong
optical input, the optical transmission coefficient can be greatly amplified
in a particular direction and suppressed in the opposite direction. The
expressions of the optimal transmission coefficient and the corresponding
isolation ratio are given analytically. Our results pave a way to design
high-quality nonreciprocal devices based on optomechanical systems.
\end{abstract}


\affiliation{$^1$ Beijing Computational Science Research Center, Beijing
100193, China}
\affiliation{$^2$ School of Mathematics, Guizhou Normal
University, Guiyang 550001, China}
\affiliation{$^3$ Department of Applied Physics, East China Jiaotong
University, Nanchang 330013, China}
\affiliation{$^4$ School of Physics and Electronic Electrical
Engineering, Huaiyin Normal University, Huai'an 223300, China}
\affiliation{$^5$ Synergetic Innovation Center
for Quantum Effects and Applications, Hunan Normal University, Changsha
410081, China}
\maketitle
\section{Introduction}

The study of optomechanical systems~\cite{aspe} based on the parametric
coupling between the photonic and phononic fields, excites a wide range of
interests. Many interesting properties of the optomechanical systems, such
as optomechanically induced transparency (OMIT)~\cite{agarwal,painter,weiss}%
, quantum entanglement~\cite{lt,ydwang}, Bell-nonlocality~\cite{bell}, and
imaging structure of tumors~\cite{tumors}, have been reported. These
properties indicate that the optomechanical system is a key quantum coherent
device for precise measurement and quantum information processing.

In a network based on electrical or optical elements, one of the key
coherent devices is the nonreciprocal one, such as isolator or circulator,
where the signals have significantly different transmission behaviors in two
opposite directions due to the breaking of time-reversal symmetry.
Traditionally, the approach to break the time-reversal symmetry is utilizing
the magneto-optical effect~\cite{MO Effect,Peterson}, which usually makes
the system bulky and unrobust to the external magnetic field. Recently,
several magnetic-free mechanisms have been proposed to implement
nonreciprocal devices, such as spatio-temporal asymmetry of refractive-index~%
\cite{ri1,ri2}, angular momentum biasing in photonic or acoustic systems~%
\cite{DLsoun13,amb1,amb2,amb3}.

As an all-optical and magnetic-free platform, the optomechanical system has
also been suggested to implement the optical nonreciprocal devices. Up to
now, there exist at least two kinds of optical nonreciprocity based on
optomechanical systems. For the first kind, the transmitted signal is the
weak light field, and its transmission behavior is assisted by another
strong control field which enhances significantly the effective
optomechanical coupling. This kind of nonreciprocity has been achieved in
physical systems displayed OMIT ~\cite{hafezi,dchomit,jh}, frequency
conversion between optical and microwave fields~\cite{tl,ok1}, and
quantum-limited amplification~\cite{clerk,
mercier,nunnenkamp,malz,fang,zxz,jc,ok,dch,zl}. And the second kind of
optical nonreciprocity is based on the nonlinear interaction in the system,
suggested in Ref.~\cite{manipat}. Here the input field (that is, the
transmitted signal) is usually very strong, and it is not necessary to
introduce the additional strong control field. A variety of nonlinear
interactions, induced by coupling the cavity fields to a qubit~\cite{zheng},
atomic ensemble~\cite{song,xia}, mechanical resonators~\cite%
{Ruesink,Rodriguez,xu}, Brillouin scattering~\cite{HF,Poulton,otterstrom},
or nonlinear optical medium~\cite{xm}, have been used to investigate this
kind of optical nonreciprocity.

\begin{figure}[tbp]
\centering
\includegraphics[width=8.5cm]{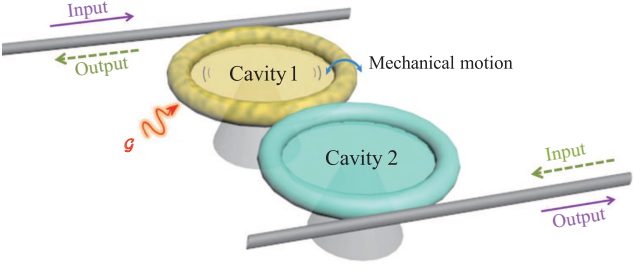} 
\caption{(Color online) Schematic diagram of the three-mode optomechanical
system with optical gain. The whispering-gallery cavity $1$ is coupled to
the mechanical mode induced by radial radiation-pressure onto the cavity
boundary~\protect\cite{kippenberg}, and the additional optical gain $%
\mathcal{G}$ is introduced for cavity ${1}$~\protect\cite{Peng,xm}. The
second whispering-gallery cavity ${2}$ is coupled to the cavity ${1}$ via
optical hopping interaction. The input field is injected either from the
cavity ${1}$ or the cavity ${2}$.}
\label{Fig-1}
\end{figure}

We would like to note that a nonreciprocal device of optical diode based on
the nonlinear interaction has recently been proposed~\cite{xu} in a
three-mode system, which is composed by a standard optomechanical system
plus another cavity coupled with the optomechanical cavity (shown in Fig.~%
\ref{Fig-1}). In this work, we will further investigate the optical
nonreciprocal phenomenon in the similar three-mode optomechanical system
with introducing an additional optical gain for the optomechanical cavity.

For the case without optical gain~\cite{xu}, the value of the transmission
coefficient is usually smaller than $1$ and the optical diode was achieved.
With the aid of the optical gain in the three-mode optomechanical system, we
find in this work that the value of the transmission coefficient in one
direction can be much larger than $1$, while in the opposite direction it
can be much smaller than $1$. Thus, the optical unidirectional amplification
can be achieved with good isolation rate due to the presence of the
additional optical gain. And the analytical expression of the optimal
transmission coefficient in the amplifying direction is obtained, which is
only determined by the product of two factors, with the first (second) term
representing the proportion of the external decay rate into the effective
(total) decay of the cavity.

\section{Model and steady-state solution}

For concreteness, the optomechanical system under consideration is
schematically shown in Fig.~\ref{Fig-1}, which consists of two coupled
whispering-gallery cavities and one mechanical resonator induced by radial
radiation-pressure onto the cavity boundary~\cite{kippenberg} of one of the
cavitis (cavity 1). In addition, the optical gain is introduced for cavity
1, which can be achieved by doped Er$^{3+}$ ions in silica with pumping the
Er$^{3+}$ ions by a laser~\cite{Peng,xm}. The Hamiltonian of such an
optomechanical system can be written as ($\hbar =1$)
\begin{eqnarray}
H &=&\omega _{1}a_{1}^{\dagger }a_{1}+\omega _{2}a_{2}^{\dagger }a_{2}+\frac{%
1}{2}\omega _{m}\left( q^{2}+p^{2}\right) +J(a_{1}^{\dagger
}a_{2}+a_{2}^{\dagger }a_{1})  \notag \\
&&+ga_{1}^{\dagger }a_{1}q+i\sqrt{\kappa _{1,e}}\left( \alpha_{1,\mathrm{in}%
}a_{1}^{\dagger }e^{-i\omega _{d}t}-\alpha_{1,\mathrm{in}}^{\ast
}a_{1}e^{i\omega _{d}t}\right)  \notag \\
&&+i\sqrt{\kappa _{2,e}}\left( \alpha_{2,\mathrm{in}}a_{2}^{\dagger
}e^{-i\omega _{d}t}-\alpha_{2,\mathrm{in}}^{\ast }a_{2}e^{i\omega
_{d}t}\right) ,  \label{H}
\end{eqnarray}%
where $a_{1}$ and $a_{2}$ are the annihilation operators of the optical
fields in two cavities (with the frequencies of $\omega _{1}$ and $\omega
_{2}$); $p$ and $q$ are the momentum and displacement operators of the
mechanical resonator (with the resonance frequency of $\omega _{m}$),
respectively. $\kappa _{j,e}$ ($j=1,2$) is the external decay rate of cavity
$j$. In Eq.~(\ref{H}), the fourth term denotes the coupling between two
cavities with strength $J$, and the fifth term represents the
radiation-pressure optomechanical coupling with the single-photon
optomechanical coupling $g$. The last two terms stand for the coupling
between the classical input fields (with the amplitude $\alpha _{j,\mathrm{in%
}}$ and the frequency of $\omega _{d}$) and the cavity fields.

According to Hamiltonian~(\ref{H}), the quantum Langevin equations (QLEs)
are obtained in the rotating frame of the driving frequency $\omega _{d}$ as
\begin{subequations}
\begin{align}
\dot{a}_{1}=& -\left( i\Delta _{1}+\frac{\kappa _{\mathrm{eff}}}{2}\right)
a_{1}-igqa_{1}-iJa_{2}+\sqrt{\kappa _{1,e}}\alpha_{1,\mathrm{in}} & &  \notag
\\
&+\sqrt{\kappa _{1,e}}a^{(e)}_{1,\mathrm{in}} +\sqrt{\kappa _{1,o}}a_{1,%
\mathrm{in}}^{\left( o\right) }+\sqrt{\mathcal{G}}a_{1,\mathrm{in}}^{\left(
\mathcal{G}\right) }\text{,} & &  \label{l1} \\
\dot{a}_{2}=& -\left( i\Delta _{2}+\frac{\kappa _{2}}{2}\right)
a_{2}-iJa_{1}+\sqrt{\kappa _{2,e}}\alpha_{2,\mathrm{in}}+\sqrt{\kappa _{2,e}}%
a^{(e)}_{2,\mathrm{in}} & &  \notag \\
&+\sqrt{\kappa _{2,o}}a_{2,\mathrm{in}}^{\left( o\right) }\text{,} & &
\label{l2} \\
\dot{q}=& \,\,\omega _{m}p\text{,} & &  \label{l3} \\
\dot{p}=& -\omega _{m}q-ga_{1}^{\dagger }a_{1}-\gamma _{m}p+\sqrt{2\gamma
_{m}}\zeta\text{,} & &  \label{l4}
\end{align}%
where $\Delta _{j}=\omega _{j}-\omega _{d}$ ($j=1,2$) is the detuning of
cavity $j$ from the input field, respectively. $\kappa _{j}=\kappa
_{j,o}+\kappa _{j,e}$ is the total decay rate of cavity $j$, where $\kappa
_{j,o}$ is the intrinsic decay rate. $\kappa _{\mathrm{eff}}=\kappa _{1}-%
\mathcal{G}$ is the effective decay rate of cavity $1$, where $\mathcal{G}$
is the gain rate induced by the doped $\mathrm{Er}^{3+}$ ions with optical
pumping. $\gamma _{m}$ is the decay rate of the mechanical resonator, $a_{j,%
\mathrm{in}}^{\left( e\right) }$, $a_{j,\mathrm{in}}^{\left( o\right) }$ $%
a_{1,\mathrm{in}}^{\left( \mathcal{G}\right) }$, and $\zeta $ are the noise
operators with zero mean values.

Assuming the input signal field(s) to be strong enough, the operators can be
replaced by their average values with the mean-field approximation $\alpha
_{j}=\left\langle a_{j}\right\rangle $, $\alpha _{j,\mathrm{in}%
}=\left\langle a_{j,\mathrm{in}}\right\rangle $, $\bar{p}=\left\langle
p\right\rangle $, and $\bar{q}=\left\langle q\right\rangle $. From Eqs.~(\ref%
{l1}-\ref{l4}), one can obtain the following steady-state equations

\end{subequations}
\begin{subequations}
\begin{align}
0=& -\left( i\Delta _{1}+\frac{\kappa _{\mathrm{eff}}}{2}\right) \alpha
_{1}-ig\bar{q}\alpha _{1}-iJ\alpha _{2}+\sqrt{\kappa _{1,e}}\alpha _{1,%
\mathrm{in}}\text{,} & &  \notag  \label{a1} \\
& & & \\
0=& -\left( i\Delta _{2}+\frac{\kappa _{2}}{2}\right) \alpha _{2}-iJ\alpha
_{1}+\sqrt{\kappa _{2,e}}\alpha _{2,\mathrm{in}}\text{,} & &  \label{a2} \\
\bar{p}=& \,\,0\text{,} & &  \label{a3} \\
\bar{q}=& -\frac{g\left\vert \alpha _{1}\right\vert ^{2}}{\omega _{m}}\text{.%
} & &  \label{a4}
\end{align}

To study the optical nonreciprocal transmission, we will focus on two cases.
In the first case, the input field is only injected into cavity ${1}$ with
amplitudes $|\alpha _{1,\mathrm{in}}|=\sqrt{p_{\mathrm{in}}/(\hbar \omega
_{d})}$ and $\alpha _{2,\mathrm{in}}=0$, where $p_{\mathrm{in}}$ is the
power of the input field. With the input-output relation~\cite{Gardiner}
\end{subequations}
\begin{equation}
\alpha _{j,\mathrm{out}}+\alpha _{j,\mathrm{in}}=\sqrt{\kappa _{j,e}}\alpha
_{j}\text{,}  \label{io}
\end{equation}%
the equation of the output field $\alpha _{2,\mathrm{out}}$ can be given as

\begin{equation}
0=-\left( \frac{\kappa }{2}+i\Delta \right) \alpha _{2,\mathrm{out}%
}+iU\left\vert \alpha _{2,\mathrm{out}}\right\vert ^{2}\alpha _{2,\mathrm{out%
}}+\varepsilon \alpha_{1,\mathrm{in}}\text{,}  \label{a1in}
\end{equation}%
where

\begin{subequations}
\begin{align}
\kappa \equiv &\,\,\kappa _{\mathrm{eff}}+\frac{4J^{2}\kappa _{2}}{\kappa
_{2}^{2}+4\Delta _{2}^{2}}\text{,} \\
\Delta \equiv &\,\,\Delta _{1}-\frac{4J^{2}\Delta _{2}}{\kappa
_{2}^{2}+4\Delta _{2}^{2}}\text{,} \\
U \equiv &\,\,\frac{g^{2}\left( \kappa _{2}^{2}+4\Delta _{2}^{2}\right) }{%
4\omega _{m}J^{2}\kappa _{2,e}}\text{,} \\
\varepsilon \equiv &-\frac{2iJ\sqrt{\kappa _{1,e}\kappa _{2,e}}}{\kappa
_{2}+2i\Delta _{2}}\text{.}
\end{align}

In the second case, the input field is only injected into cavity ${2}$ with
the amplitude $|\tilde{\alpha}_{2,\mathrm{in}}|=\sqrt{\tilde{p}_{\mathrm{in}%
}/(\hbar \omega _{d})}$ and $\tilde{\alpha}_{1,\mathrm{in}}=0$, where $%
\tilde{p}_{\mathrm{in}}$ is the power of input field. Here we have added
tildes \textquotedblleft $\tilde{\text{ }}$\textquotedblright\ for ${\alpha }%
_{j,\mathrm{in}}$, ${\alpha }_{j,\mathrm{out}}$, and ${p}_{\mathrm{in}}$ in
order to distinguish them from that in the first case.

Similarly, the equation of the output field $\tilde{\alpha}_{1,\mathrm{out}}$
is obtained as

\end{subequations}
\begin{equation}
0=-\left( \frac{\kappa }{2}+i\Delta \right) \tilde{\alpha}_{1,\mathrm{out}%
}+i \tilde{U}\left\vert \tilde{\alpha}_{1,\mathrm{out}}\right\vert ^{2}%
\tilde{\alpha}_{1,\mathrm{out}}+{\varepsilon} \tilde{\alpha}_{2,\mathrm{in}}
\text{,}  \label{a2in}
\end{equation}%
where
\begin{equation}
\tilde{U}\equiv \frac{g^{2}}{\omega _{m}\kappa_{1,e}}.
\end{equation}

To describe the transmission properties quantitatively, we define the
following transmission coefficients
\begin{equation}
T\equiv \left\vert \frac{\alpha _{2,\mathrm{out}}}{\alpha _{1,\mathrm{in}}}%
\right\vert ^{2}\text{,}\ \ \tilde{T}\equiv \left\vert \frac{\tilde{\alpha}%
_{1,\mathrm{out}}}{\tilde{\alpha}_{2,\mathrm{in}}}\right\vert ^{2},
\end{equation}%
respectively, for the two cases with opposite transmission directions.

By making use of Eqs.~(\ref{a1in}) and (\ref{a2in}), the transmission
coefficients are determined by
\begin{subequations}
\begin{align}
0 =&\,\,4U^{2}T^{3}s_{\mathrm{in}}^{2}-8\Delta UT^{2}s_{\mathrm{in}}+T\left(
\kappa ^{2}+4\Delta ^{2}\right) -\lambda \text{,} &   \notag \\
& & \label{t21}  \\
0 =&\,\,4\tilde{U}^{2}\tilde{T}^{3}\tilde{s}_{\mathrm{in}}^{2}-8\Delta
\tilde{U}\tilde{T}^{2}\tilde{s}_{\mathrm{in}}+\tilde{T}\left( \kappa
^{2}+4\Delta ^{2}\right) -\lambda &   \notag \\
& & \label{t12}
\end{align}
with $s_{\mathrm{in}}=|\alpha_{1,\mathrm{in}}|^{2}$, $\tilde{s}_{\mathrm{in}%
}=|\alpha_{2,\mathrm{in}}|^{2}$, and $\lambda =16J^{2}\kappa _{1,e}\kappa
_{2,e}/\left( \kappa_{2}^{2}+4\Delta _{2}^{2}\right) $.

The optical nonreciprocity requires $T\neq \tilde{T}$ when the input fields
have the same powers in the two cases, i.e. $p_{\mathrm{in}}=\tilde{p}_{%
\mathrm{in}}$ and $s_{\mathrm{in}}=\tilde{s}_{\mathrm{in}}$. Thus it is
clear from Eqs.~(\ref{t21}) and (\ref{t12}) that the necessary condition to
observe the optical nonreciprocity is $U\neq \tilde{U}$, which can be
explicitly written as
\end{subequations}
\begin{equation}
\kappa _{1,e}\left( \kappa _{2}^{2}+4\Delta _{2}^{2}\right) \neq 4\kappa
_{2,e}J^{2}\text{.}  \label{non}
\end{equation}

We would like to note that the similar condition of optical nonreciprocity
has also been reported in Ref.~\cite{xu} in a similar three-mode
opotomechanical system without the optical gain.

\begin{figure*}[th]
\centering
\subfigure{
\includegraphics[width=5.5cm]{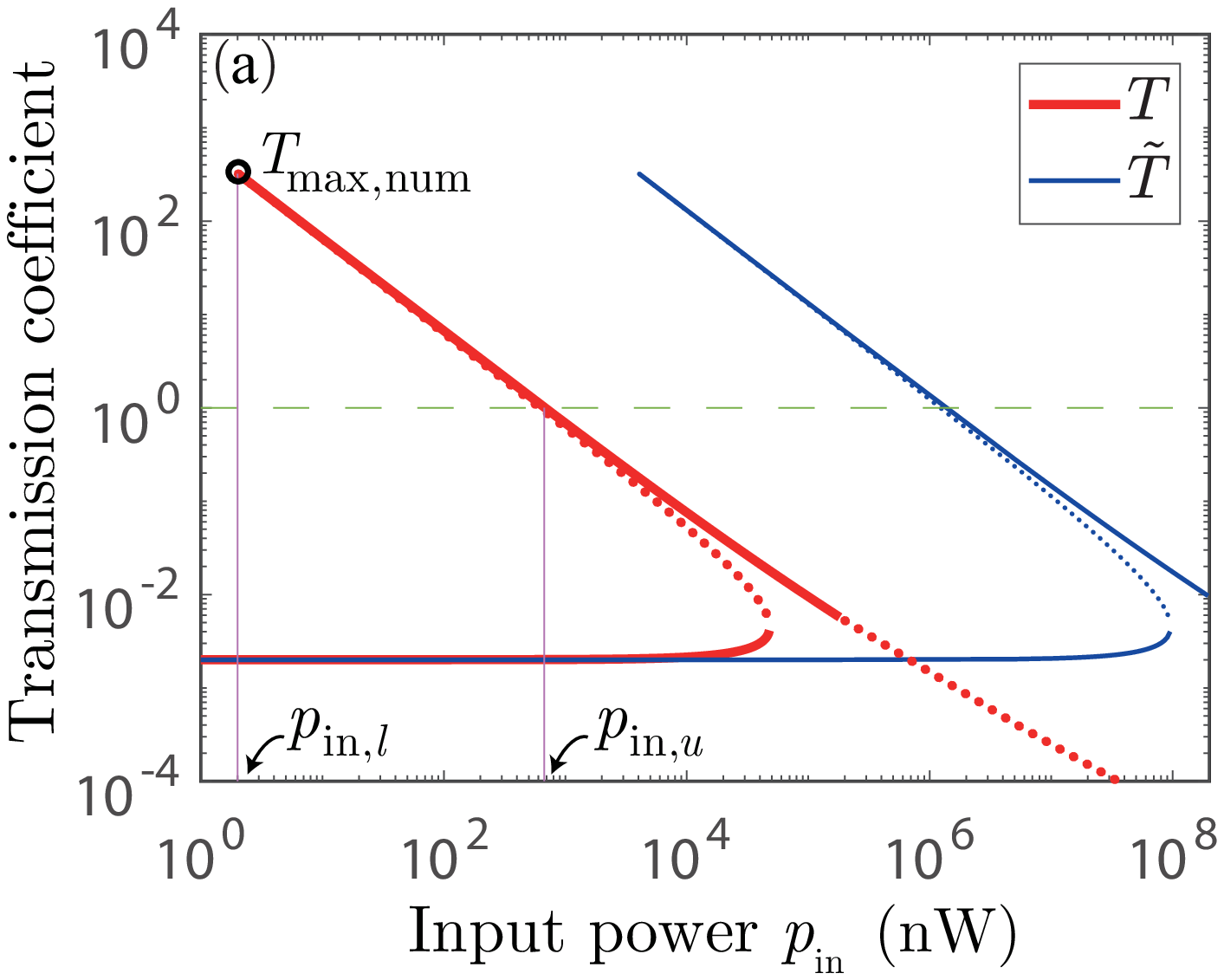} }
\subfigure
{\includegraphics[width=5.5cm]{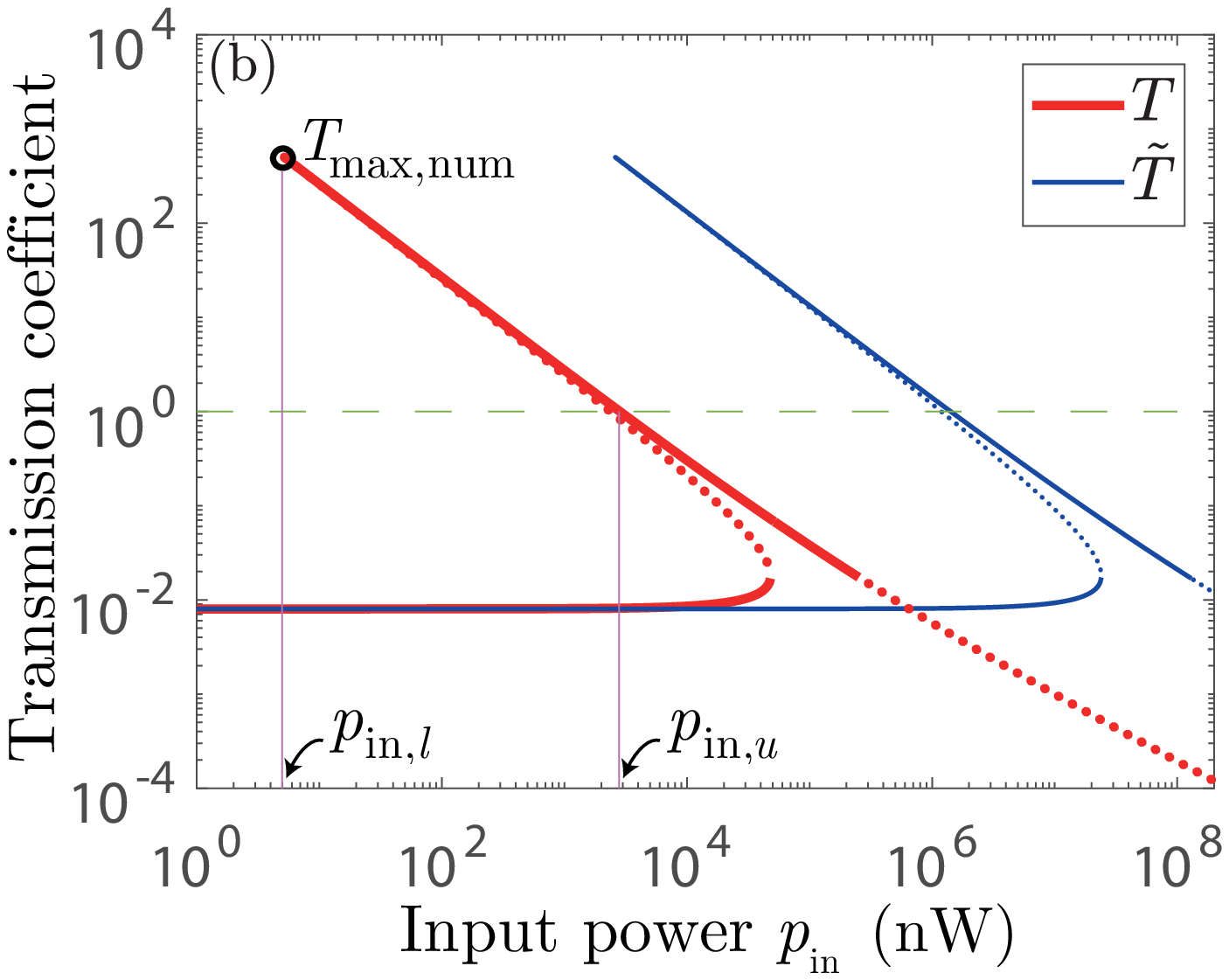}} \subfigure
{\includegraphics[width=5.5cm]{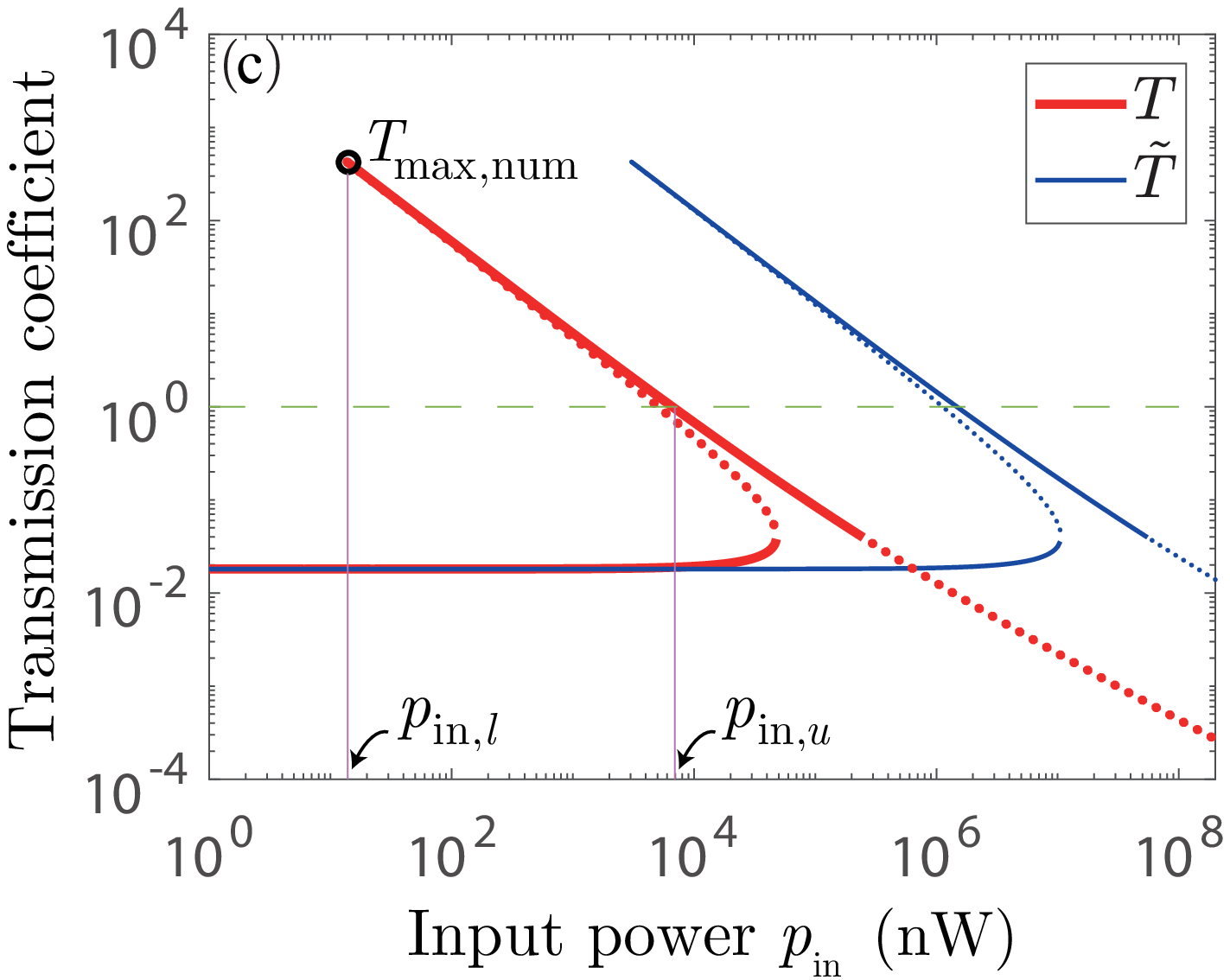} }
\caption{(Color online) Transmission coefficients $T$ (red) and $%
\tilde{T}$ (blue) as a function of the input power $p_{\mathrm{in}}$
for (a) $J=0.5J_{0} $, (b) $J=J_{0}$, and (c) $J=1.5J_{0}$. $T_{\mathrm{%
max,num}}$ (black circle) represents the maximum value of the transmission
coefficient $T$ obtained numerically. The solid (dotted) lines represent
stable (unstable) values of $T$ and $\tilde{T}$. The green dash line
corresponding to $T=1$ is the benchmark of amplification. The system shows
clearly the working region of optical nonreciprocity is $p_{\mathrm{in}}\in
[p_{\mathrm{in},l},~p_{\mathrm{in},u}]$. Here the other parameters are
chosen referring to the recent optomechanical experiment with whispering
gallery~\protect\cite{par}: $\protect\kappa _{1}/2\protect\pi = \protect%
\kappa _{2}/2\protect\pi =100\,\mathrm{MHz}$, $\protect\kappa _{\mathrm{eff}%
}/2\protect\pi =200\,\mathrm{kHz}$, $\protect\kappa _{1,e}/2\protect\pi =%
\protect\kappa _{2,e}/2\protect\pi =100\,\mathrm{MHz}$, $\protect\omega %
_{d}/2\protect\pi =200\,\mathrm{THz}$, $\protect\omega _{m}/2\protect\pi %
=200\,\mathrm{MHz}$, $\protect\gamma _{m}/2\protect\pi =50\,\mathrm{kHz}$, $%
g/2\protect\pi =0.8\,\mathrm{kHz}$, $\Delta _{1}/2\protect\pi =50\,\mathrm{%
MHz}$, $\Delta _{2}/2\protect\pi =20\,\mathrm{MHz}$, and $J_{0}/2\protect\pi %
=2.41\,\mathrm{MHz}$. }
\label{Fig-2}
\end{figure*}

\section{Unidirectional amplification}

In this section we will study the transmission behavior in the three-mode
optomechanical system under consideration. It is found that the optical
signal field can be unidirectionally amplified with the additional optical
gain. And the expressions of the optimal transmission coefficient and the
isolation ratio are given analytically.

\subsection{Stability condition}

Since both the optical gain and the nonlinear interaction are introduced in
our system, the first step is to ensure the stability of the system in
steady state. By splitting each operator into its mean value and
fluctuation: $a_{j}=\alpha _{j}+\delta a_{j}$, $q=\bar{q}+\delta q$, $p=\bar{%
p}+\delta p$, the linearized QLEs corresponding to Eqs.~(\ref{l1})-(\ref{l4}%
) can be written in a matrix form as
\begin{equation}
\dot{\mu}=-M\mu +\Gamma\mu _{\mathrm{in}}\text{,}  \label{muuu}
\end{equation}%
where $\mu =(\delta a_{1}$, $\delta a_{1}^{\dagger }$, $\delta a_{2}$, $%
\delta a_{2}^{\dagger }$, $\delta p$, $\delta q)^{T}$, $\mu _{\mathrm{in}%
}=(a^{(e)}_{1,\mathrm{in}}$, $a^{(e)\dagger}_{1,\mathrm{in}}$, $a_{1,\mathrm{%
in}}^{\left( o\right) }$, $a_{1,\mathrm{in}}^{\left( o\right) \dagger }$, $%
a^{(e)}_{2,\mathrm{in}}$, $a_{2,\mathrm{in}}^{(e)\dagger }$, $a_{2,\mathrm{in%
}}^{\left( o\right) }$, $a_{2,\mathrm{in}}^{\left( o\right) \dagger }$, $%
a_{1,\mathrm{in}}^{\left( \mathcal{G}\right) }$, $a_{1,\mathrm{in}}^{\left(
\mathcal{G}\right) \dagger }$, $0$, $\zeta )^{T}$, and the coefficient matrix

\begin{widetext}
\begin{equation}
M=\left(
\begin{array}{cccccc}
\frac{\kappa _{\mathrm{eff}}}{2}+i\left( \Delta _{1}+g\bar{q}
\right)  & 0 & iJ & 0 & ig\alpha _{1} & 0 \\
0 &  \frac{\kappa _{\mathrm{eff}}}{2}-i\left( \Delta _{1}+g\bar{q}\right)
  & 0 & -iJ & -ig\alpha _{1}^{\ast } & 0 \\
iJ & 0 &  \frac{\kappa _{2}}{2}+i\Delta _{2}  & 0 & 0 & 0 \\
0 & -iJ & 0 & \frac{\kappa _{2}}{2}-i\Delta _{2}  & 0 & 0 \\
0 & 0 & 0 & 0 & 0 & -\omega _{m} \\
g\alpha _{1}^{\ast } & g\alpha _{1} & 0 & 0 & \omega _{m} & \gamma _{m}%
\end{array}%
\right) \text{,} \label{m}
\end{equation}%
\begin{equation}
\Gamma=\left(
\begin{array}{cccccccccccc}
\sqrt{\kappa
_{1,e}} & 0 & \sqrt{\kappa _{1,o}} & 0 & 0 & 0 & 0 & 0 & \sqrt{\mathcal{G}}  & 0 & 0 & 0 \\%
 0 & \sqrt{\kappa _{1,e}} & 0 & \sqrt{\kappa _{1,o}} & 0 & 0 & 0 & 0 & 0 & \sqrt{\mathcal{G}} & 0 & 0 \\%
  0 & 0 & 0 & 0 &
\sqrt{\kappa _{2,e}} & 0 & \sqrt{\kappa _{2,o}} & 0 & 0 & 0 & 0 & 0 \\ 0 & 0 & 0 & 0 & 0 & \sqrt{\kappa _{2,e}} & 0 & \sqrt{\kappa _{2,o}} & 0 & 0 & 0 & 0 \\ %
0 & 0 & 0 & 0 & 0 & 0 & 0 & 0 & 0 & 0 & 0 & 0 \\ 0 & 0 & 0 & 0 & 0 & 0 & 0 & 0 & 0 & 0 & 0 & \sqrt{2\gamma _{m}}
\end{array}%
\right) \text{.}
\end{equation}
\end{widetext}
The stability condition can be derived by using the Routh-Hurwitz criterion~%
\cite{dejusus}, which requires all the real parts of eigenvalues of the
matrix $M$ to be positive. The explicit forms of such a criterion in the
current model are cumbersome and not given here. However, in the following
discussions all the stability conditions have been checked numerically.

\subsection{Optical amplification induced by optical gain}

For the nonreciprocal device based on the nonlinearity in the three-mode
optomechanical system~\cite{xu}, the optical diode is achieved and the value
of the maximum transmission coefficients is usually smaller than one. This
subsection will show the optical unidirectional amplification assisted by
the optical gain. That is, the transmission coefficient along one of the two
directions is larger than one, and the one in the opposite direction is much
smaller than one.

In Fig.~\ref{Fig-2}, the transmission coefficients $T$ and $\tilde{T}$ are
plotted as a function of the input power $p_{\mathrm{in}}$. It is apparent
that in Fig.~\ref{Fig-2} the optical unidirectional amplification appears in
two regions where $T>1>\tilde{T}$ (i.e. $p_{\mathrm{in}}\in \lbrack p_{%
\mathrm{in},l},p_{\mathrm{in},u}]$) and $\tilde{T}>1>T$, respectively.
However, the isolation ratio in the first region is better than that in the
second region. Then in what follows, we just focus on the first region with $%
p_{\mathrm{in}}\in \lbrack p_{\mathrm{in},l},p_{\mathrm{in},u}]$, where we
only consider the upper branch of $T$.

As shown in Fig.~\ref{Fig-2}(a), when the system works in the upper branch
of $T$ with $p_{\mathrm{in}}\in \lbrack p_{\mathrm{in},l},~p_{\mathrm{in}%
,u}] $, it has obvious optical nonreciprocity with the tremendous difference
between the values of (upper-branch) $T$ and $\tilde{T}$. Here, $p_{\mathrm{%
in},l}=2.03\,\mathrm{nW}$ and $p_{\mathrm{in},u}=0.68\,\mathrm{{\mu}W}$
corresponding to $T=T_{\mathrm{max,num}}$ and $T=1$, respectively, are the
lower and upper bounds of input field power.

\begin{figure*}[th]
\centering
\subfigure{
\includegraphics[width=5.5cm]{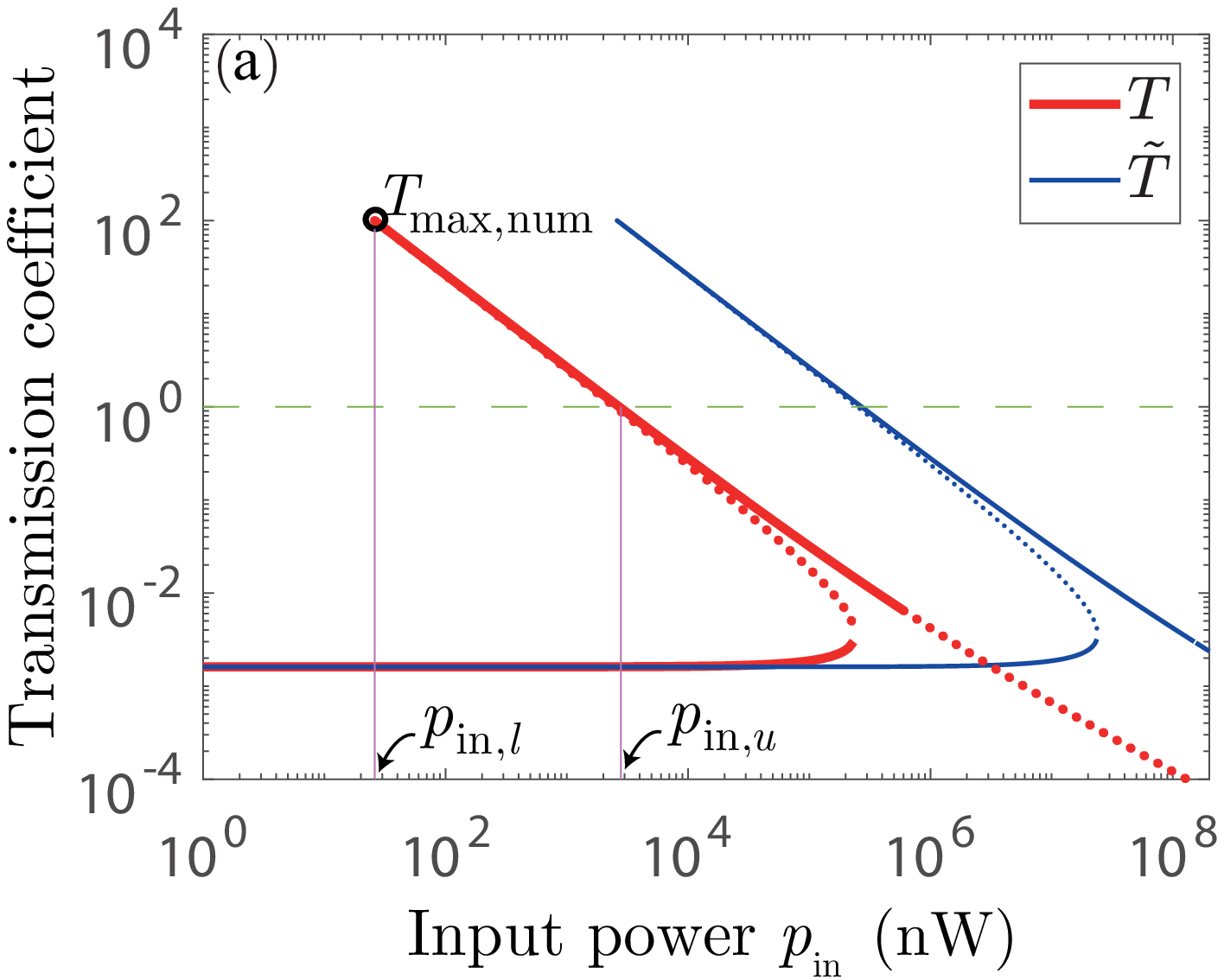} }
\subfigure{
\includegraphics[width=5.5cm]{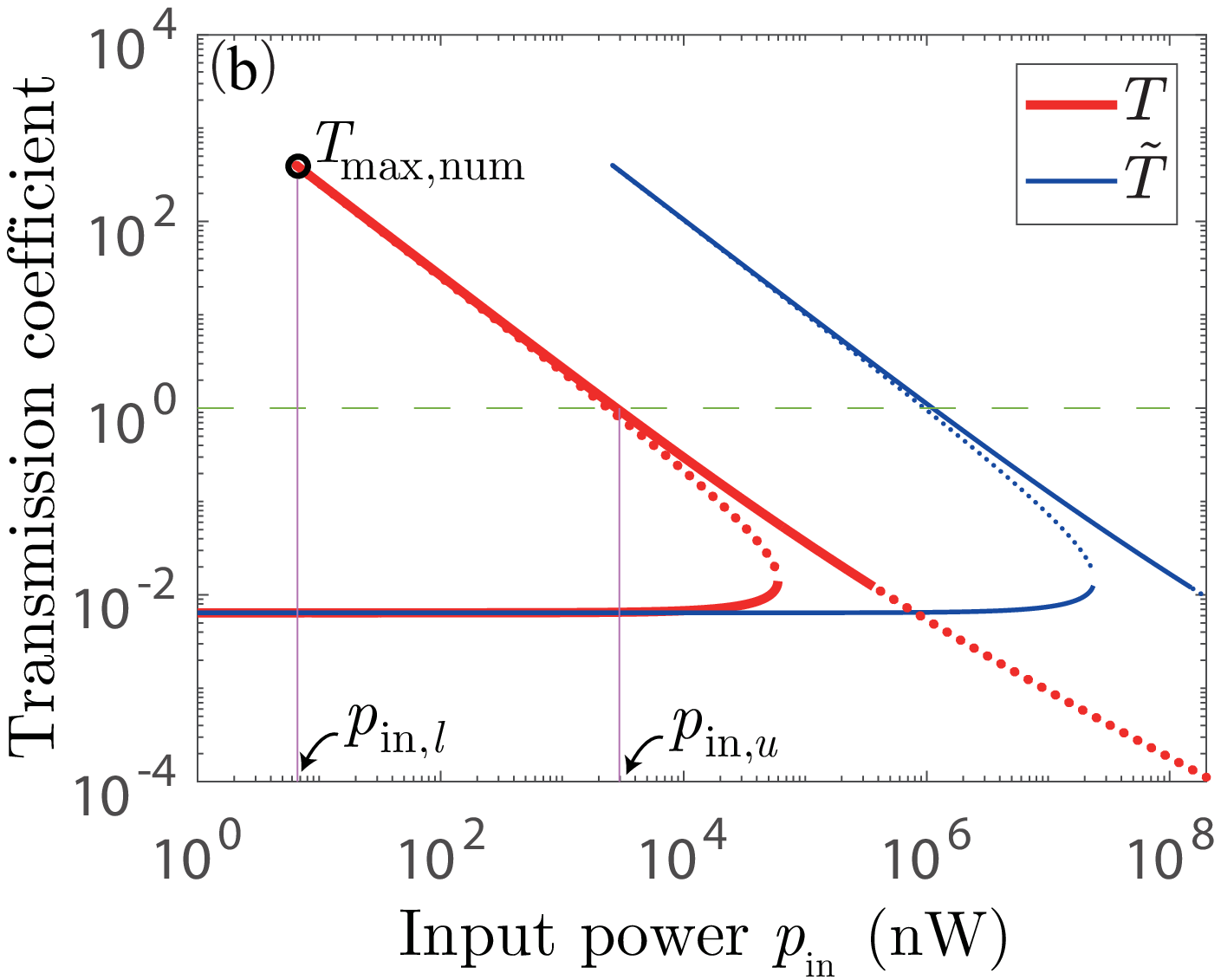} } \subfigure
{\includegraphics[width=5.5cm]{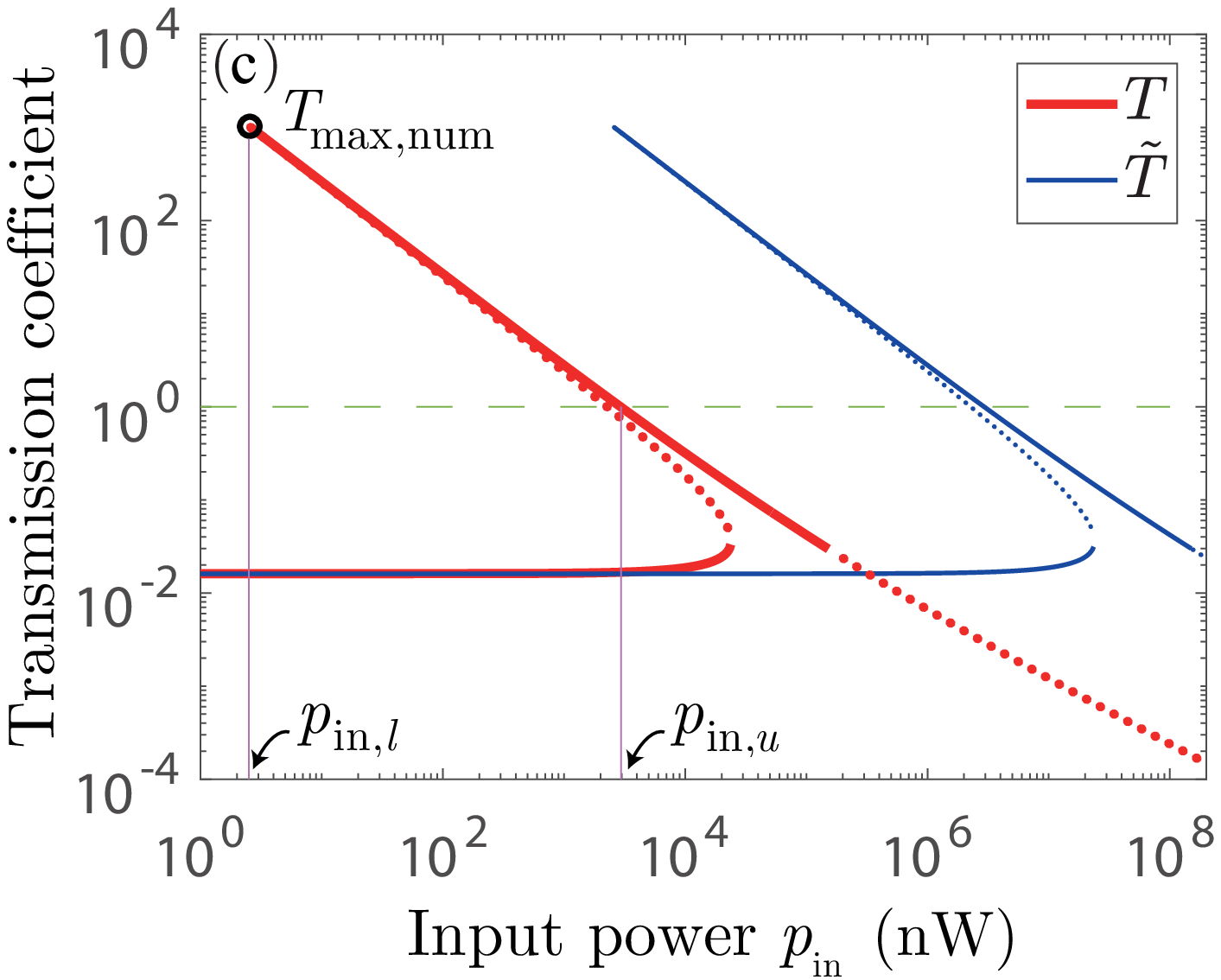}}
\caption{(Color online) Transmission coefficients $T$ (red) and $\tilde{T}$
(blue) as a function of the input power $p_{\mathrm{in}}$ for (a) $\protect%
\kappa _{1,e}/2\protect\pi =20\,\mathrm{MHz}$, (b) $\protect\kappa _{1,e}/2%
\protect\pi =80\,\mathrm{MHz}$, and (c) $\protect\kappa _{1,e}/2\protect\pi %
=200\,\mathrm{MHz}$. $T_{\mathrm{max,num}}$ (the black circle) represents
the maximum value of the transmission coefficient $T$ obtained numerically.
Here the other parameters are chosen referring to Ref.~\protect\cite{par}: $%
\protect\kappa _{1}/2\protect\pi =200\,\mathrm{MHz}$, $\protect\kappa _{2}/2%
\protect\pi =100\,\mathrm{MHz}$, $\protect\kappa _{\mathrm{eff}}/2\protect%
\pi =200\,\mathrm{kHz}$, $\protect\kappa _{2,e}/2\protect\pi =100\,\mathrm{%
MHz}$, $\protect\omega _{d}/2\protect\pi =200\,\mathrm{THz}$, $\protect%
\omega _{m}/2\protect\pi =200\,\mathrm{MHz}$, $\protect\gamma _{m}/2\protect%
\pi =50\,\mathrm{kHz}$, $g/2\protect\pi =0.8\,\mathrm{kHz}$, $\Delta _{1}/2%
\protect\pi =50\,\mathrm{MHz}$, $J/2\protect\pi=2.19\,\mathrm{MHz}$, $\Delta
_{2}/2\protect\pi =60\,\mathrm{MHz}$. Similar to Fig.~\protect\ref{Fig-2},
the solid (dotted) lines represent the stable (unstable) values for $T$ and $%
\tilde T$, and the green dash line corresponding to $T=1$ is the benchmark
of amplification. }
\label{Fig-5}
\end{figure*}

To quantify optical nonreciprocity, the isolation ratio is introduced as $%
E\left( \mathrm{dB}\right) =10\times \log _{10}({T}/\tilde{T})$.
Accordingly, with $p_{\mathrm{in}}\in \lbrack p_{\mathrm{in},l},p_{\mathrm{in%
},u}]$ in Fig.~\ref{Fig-2}(a), it is found numerically that $|E\left(
\mathrm{dB}\right) |\in \lbrack 26.99,52.04]$. Moreover, in Fig.~\ref{Fig-2}%
(a) the value of $T$ is larger than $1$ while that of $\tilde{T}$ is much
smaller than $1$ in the working region. It clearly displays that the signal
is amplified when the input field is injected from cavity $1$. With the aid
of the optical gain, Fig.~\ref{Fig-2}(b) and Fig.~\ref{Fig-2}(c) also show
the similar unidirectional amplification as that in Fig.~\ref{Fig-2}(a).
Note that in Fig.~\ref{Fig-2}, the parameters satisfy the nonreciprocity
condition Eq.~(\ref{non}).

The effect of the external decay rate $\kappa _{1,e}$ on the transmission
behavior is also investigated in Fig.~\ref{Fig-5}. This figure shows that
with the increase of $\kappa _{1,e}$, all the values of the transmission
coefficients are collectively lifted upward. This means with the increase of
the external decay in cavity ${1}$, both the transmission coefficients in
the two directions can be increased with the unidirectional amplification
remained.

\begin{figure}[tbp]
\centering
\includegraphics[width=8cm]{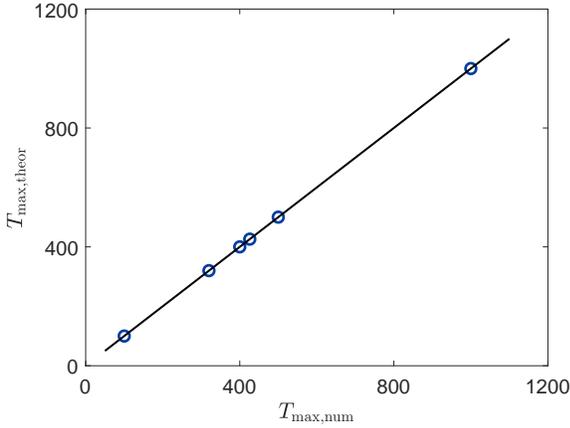} 
\caption{(Color online) The black solid line represents the equation $T_{%
\mathrm{max}, \mathrm{num}}=T_{\mathrm{max}, \mathrm{theor}}$, and each blue
circle denotes the point with the coordinates ($T_{\mathrm{max},\mathrm{num}%
} $, $T_{\mathrm{max}, \mathrm{theor}}$) which are respectively obtained
numerically and analytically for the same parameters. Actually these values
of $T_{\mathrm{max},\mathrm{num}} $ are taken from the data marked as black
circles in Fig.~\protect\ref{Fig-2} and Fig.~\protect\ref{Fig-5}, and that
of $T_{\mathrm{max}, \mathrm{theor}}$ are calculated based on Eq.~(\protect
\ref{tup}) with the corresponding parameters. One can find all the blue
circles collapse into the line.}
\label{Fig-3}
\end{figure}

\begin{figure}[tbp]
\centering
\includegraphics[width=8cm]{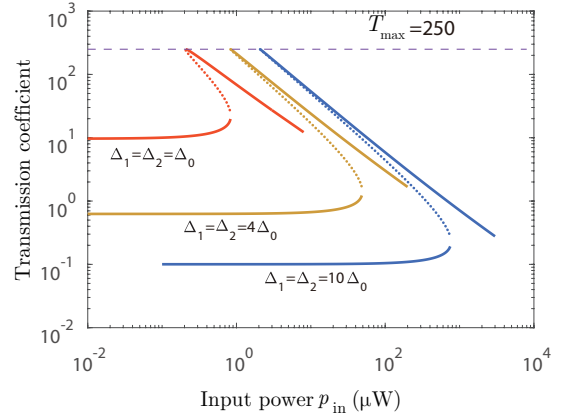} 
\caption{(Color online) Transmission coefficients $T$ as a function of the
input power $p_{\mathrm{in}}$ for different values of the detuning. The
dotted lines represent unstable values and solid lines represent stable
values. The cavity coupling strength is kept to be the optimal one $J=J_{%
\text{opt}}\equiv\protect\sqrt{\protect\kappa _{\mathrm{eff}}(\protect\kappa %
_{2}^{2}+\Delta _{2}^{2})/(4\protect\kappa _{2})}$. Here the other
parameters are chosen based on a recent optomechanical experiment with
whispering gallery~\protect\cite{par}: $\protect\kappa _{1}/2\protect\pi %
=50\,\mathrm{MHz}$, $\protect\kappa _{2}/2\protect\pi =100\,\mathrm{MHz}$, $%
\protect\kappa _{\mathrm{eff}}/2\protect\pi =200\,\mathrm{kHz}$, $\protect%
\kappa _{1,e}/2\protect\pi =50\,\mathrm{MHz}$, $\protect\kappa _{2,e}/2%
\protect\pi =100\,\mathrm{MHz}$, $\protect\omega _{d}/2\protect\pi =200\,%
\mathrm{THz}$, $\protect\omega _{m}/2\protect\pi =200\,\mathrm{MHz}$, $%
\protect\gamma _{m}/2\protect\pi =50\,\mathrm{kHz}$, $g/2\protect\pi =0.8\,%
\mathrm{kHz}$, $\Delta _{0}/2\protect\pi =1\,\mathrm{MHz}$.}
\label{Fig-4}
\end{figure}

\subsection{Optimal transmission coefficient and the corresponding isolation
ratio}

In Sec.~III B, it is found that with $p_{\mathrm{in}}\in \lbrack p_{\mathrm{%
in},l},~p_{\mathrm{in},u}]$, our optomechanical system displays the optical
nonreciprocal transmission of unidirectional amplification. This inspires us
to ask the following question: What are the optimal maximum transmission
coefficient and the corresponding isolation ratio in our system? We will
study such a question in details in this section.

Eq.~(\ref{t21}) is a cubic equation for the transmission coefficient $T$.
However, the analytical solution of $T$ has somewhat complex dependence on
the system parameters and makes it less informative. This difficulty can be
circumvented by solving $s_{\mathrm{in}}$ in Eq.~(\ref{t21}). The solution
to $s_{\mathrm{in}}$ in Eq.~(\ref{t21}) is formally given as
\begin{equation}
s_{\mathrm{in}}=\frac{2T\Delta \pm \sqrt{T\lambda -T^{2}\kappa ^{2}}}{2T^{2}U%
}\text{.}  \label{s1in}
\end{equation}%
Because $s_{\mathrm{in}}$ must be positive, under the condition $\Delta >0$,
the valid region of $T$ with $T\lambda -T^{2}\kappa ^{2}\geq 0$ should be
\begin{equation}
0<T\leq T_{\mathrm{max}\mathrm{,theor}}\text{,}  \label{tr21}
\end{equation}%
where the possible maximum transmission coefficient
\begin{equation}
T_{\mathrm{max}\mathrm{,theor}}=\frac{\lambda }{\kappa ^{2}}=\frac{%
16J^{2}\kappa _{1,e}\kappa _{2,e}\left( \kappa _{2}^{2}+4\Delta
_{2}^{2}\right) }{\left[ 4J^{2}\kappa _{2}+\kappa _{\mathrm{eff}}\left(
\kappa _{2}^{2}+4\Delta _{2}^{2}\right) \right] ^{2}}\text{.}  \label{tup}
\end{equation}%
With the optical amplification requirement $T_{\mathrm{max}\mathrm{,theor}%
}>1 $, the condition for $\kappa _{\mathrm{eff}}$ is determined as
\begin{equation}
0<\kappa _{\mathrm{eff}}<\sqrt{16J^{2}\kappa _{1,e}\kappa _{2,e}\left(
\kappa _{2}^{2}+4\Delta _{2}^{2}\right) }-\frac{4J^{2}\kappa _{2}}{\kappa
_{2}+4\Delta _{2}^{2}}\text{.}  \label{keff}
\end{equation}

The numerical counterpart $T_{\mathrm{max},\mathrm{num}}$ of the maximum
transmission coefficient $T_{\mathrm{max}}$ can be easily obtained by the
numerical solutions to Eq.~(\ref{t21}), such as that in Fig.~\ref{Fig-2}.
For the parameters considered in Figs.~\ref{Fig-2}-\ref{Fig-3}, it is
checked that the relation $T_{\mathrm{max},\mathrm{num}}=T_{\mathrm{max}%
\mathrm{,theor}}$ is always valid in the working region. As an example, in
Fig.~\ref{Fig-3} all the blue circles representing the point ($T_{\mathrm{max%
}\mathrm{,num}}$, $T_{\mathrm{max}\mathrm{,theor}}$) collapse into the line
with equation $T_{\mathrm{max}\mathrm{,num}}=T_{\mathrm{max}\mathrm{,theor}}$%
. This suggests that the expression given in Eq.~(\ref{tup}) is a good
approximate result for $T_{\mathrm{max}}$ for the parameters considered in
Figs.~\ref{Fig-2}-\ref{Fig-3}. From now on, for simplicity we set $T_{%
\mathrm{max}}=T_{\mathrm{max}\mathrm{,theor}}$.

Then, $T_{\mathrm{max}}$ can be further optimized with respect to the
coupling strength $J$ between the two cavities. Solving $\partial T_{\mathrm{%
max}}/\partial J=0$ under the condition $\kappa _{\mathrm{eff}}>0$, the
optimal coupling strength is given as
\begin{equation}
J=J_{\mathrm{opt}}:=\sqrt{\frac{\kappa _{\mathrm{eff}}\left( \kappa
_{2}^{2}+4\Delta _{2}^{2}\right) }{4\kappa _{2}}}\text{.}  \label{jopt}
\end{equation}%
Substituting Eq.~(\ref{jopt}) into Eq.~(\ref{tup}), the optimized value of $%
T_{\mathrm{max}}$ is obtained as
\begin{equation}
T_{\mathrm{max}}^{\mathrm{opt}}=\frac{\kappa _{1,e}}{\kappa _{\mathrm{eff}}}%
\cdot \frac{\kappa _{2,e}}{\kappa _{2}}\text{.}  \label{tupopt}
\end{equation}

There are two terms in Eq.~(\ref{tupopt}), in which the first (second) term
represents the proportion of the external decay rate into the effective
(total) decay of the cavity. This indicates that $T_{\mathrm{max}}^{\mathrm{%
opt}}$ is determined only by the intrinsic parameters of the system. As a
result, $T_{\mathrm{max}}^{\mathrm{opt}}$ should remain as a constant, when
the other parameters (e.g., the detunings) are changed. This invariance of $%
T_{\mathrm{max}}^{\mathrm{opt}}$ is displayed in Fig.~\ref{Fig-4}: although
the detuning $\Delta _{1}$ and $\Delta _{2}$ change, $T_{\mathrm{max}}^{%
\mathrm{opt}}$ is unaltered.

Finally, the isolation ratio $E_{0}$ corresponding to $T_{\mathrm{max}}^{%
\mathrm{opt}}$ is derived. According to Eqs.~(\ref{t21},\ref{t12},\ref{jopt}%
), the absolute value of isolation ratio is given as
\begin{equation}
|E_{0}|\simeq 10\times \log _{10}\left( 1+\frac{\left( \kappa _{2}\Delta
_{1}-\kappa _{\mathrm{eff}}\Delta _{2}\right) ^{2}}{\kappa _{2}^{2}\kappa _{%
\mathrm{eff}}^{2}}\right),  \label{E}
\end{equation}
where we have used the fact that $T_{\mathrm{max}}^{\mathrm{opt}} \gg 1$ and
the corresponding value of $\tilde T$ at $p_{\mathrm{in}}=p_{\mathrm{in},l}$
is much less than 1. For the special case $\kappa _{2}\gg \kappa _{\mathrm{%
eff}}$ and $\Delta _{1}\sim \Delta _{{2}}\gg \kappa _{\mathrm{eff}}>0$, Eq.~(%
\ref{E}) is simplified as
\begin{equation}
|E_{0}|\simeq 10\times \log _{10} \frac{\Delta^2 _{1}}{\kappa^2 _{\mathrm{eff%
}}} .  \label{E2}
\end{equation}%
That means one can obtain good isolation ratio by modifying the optical gain
so that the effective decay rate $\kappa_{\mathrm{eff}}$ of cavity 1 is very
small compared with $\kappa_2$ and $\Delta_{1,2}$.

\section{Noise analysis}

In this section, we will analyze the effect of the added noise in our proposal.
For this, we resort to the linearized QLEs of operator fluctuations [i.e., Eq.~(\ref{muuu})],
which include the noise operators. In both cases that the input field is only injected into cavity 1
or cavity 2, Eq.~(\ref{muuu}) maintains the same expression except that the average values [e.g., $\alpha_{1}$, $\alpha_{2}$ and $\bar{q}$ in Eq.~(\ref{m})] are different in different cases.

The solution to Eq.~(\ref{muuu}) in the frequency domain can be written as
\begin{equation}
\mu \left( \omega \right) =\left( M-i\omega I\right) ^{-1}\Gamma \mu _{%
\mathrm{in}}\left( \omega \right) \text{,}  \label{muf}
\end{equation}
and the Fourier transform of any operator is introduced as
\begin{eqnarray}
o\left( \omega \right) &=&\int_{-\infty }^{+\infty }o\left( t\right)
e^{i\omega t}dt\text{.}  \label{ft}
\end{eqnarray}
Then taking Eq.~(\ref{muf}) into Eq.~(\ref{io}) in the Fourier domain, we
obtain
\begin{equation}
\mu _{\mathrm{out}}\left( \omega \right) =\mathscr{T}\left( \omega \right)
\mu _{\mathrm{in}}\left( \omega \right) \text{,}
\end{equation}
where $\mu _{\mathrm{out}}=(a^{(e)}_{1,\mathrm{out}}$, $a^{(e)\dagger}_{1,%
\mathrm{out}}$, $a_{1,\mathrm{out}}^{\left( o\right) }$, $a_{1,\mathrm{out}%
}^{\left( o\right) \dagger }$, $a^{(e)}_{2,\mathrm{out}}$, $a_{2,\mathrm{out}%
}^{(e)\dagger }$, $a_{2,\mathrm{out}}^{\left( o\right) }$, $a_{2,\mathrm{out}%
}^{\left( o\right) \dagger }$, $a_{1,\mathrm{out}}^{\left( \mathcal{G}%
\right) }$, $a_{1,\mathrm{out}}^{\left( \mathcal{G}\right) \dagger }$, $0$, $%
\zeta)^{T}$, and the scattering matrix is
\begin{equation}
\mathscr{T}\left( \omega \right) =\Gamma ^{T}\left( M-i\omega I\right)
^{-1}\Gamma -I\text{.}
\end{equation}
The element of the scattering matrix $\mathscr{T}_{ij}$ ($i,j=1,2,...,7$)
represents the transmission amplitude of the $j$th element in $\mu _{\mathrm{in}%
}\left( \omega \right)$ to the $i$th element in $\mu _{\mathrm{out}}\left(
\omega \right)$.

To calculate the output spectra, we use the non-zero
correlation functions of the input noise operators in Eq.~(\ref{muf}) as the
followings
\begin{subequations}
\begin{eqnarray}
\left\langle a_{j,\mathrm{in}}^{(e)}\left( \omega\right) a_{j,\mathrm{in}%
}^{(e)\dagger }\left( \omega^{\prime }\right) \right\rangle &=&2\pi\delta
\left( \omega+\omega^{\prime }\right) \text{, } \\
\left\langle a_{j,\mathrm{in}}^{\left( o\right) }\left( \omega\right) a_{j,%
\mathrm{in}}^{\left( o\right) \dagger }\left( \omega^{\prime }\right)
\right\rangle &=&2\pi\delta \left( \omega+\omega^{\prime }\right) \text{, }
\\
\left\langle a_{1,\mathrm{in}}^{\left( \mathcal{G}\right) \dagger }\left(
\omega\right) a_{1,\mathrm{in}}^{\left( \mathcal{G}\right) }\left(
\omega^{\prime }\right) \right\rangle &=&2\pi\delta \left(
\omega+\omega^{\prime }\right) \text{, } \\
\left\langle \zeta \left( \omega\right) \zeta \left( \omega^{\prime }\right)
\right\rangle &=&2\pi(n_{m}+\frac{1}{2})\delta \left(
\omega+\omega^{\prime }\right) \text{.}  \label{corf}
\end{eqnarray}%
\end{subequations}
Here, the thermal photon numbers have been taken to be zero as the frequencies of the cavities are very high
(e.g., of the order of $10^{14}\,\mathrm{Hz}$), however the thermal phonon number is given as $%
n_{m}=1/[\exp (\hbar \omega _{m}/k_{B}T)-1]$, where $k_{B}$ is the Boltzmann
constant and $T$ is the effective temperature of the reservoir .

Then the output spectra of cavity 2 in the first case, where the input filed
is only injected into cavity 1, can be obtained as~\cite{clerk}
\begin{widetext}
\begin{eqnarray}
S_{2,\mathrm{out}}\left( \omega \right)  &=&\frac{1}{2}\int \frac{d\omega
^{\prime }}{2\pi }\left\langle a_{2,\mathrm{out}}\left( \omega \right) a_{2,%
\mathrm{out}}^{\dagger }\left( \omega ^{\prime }\right) +a_{2,\mathrm{out}%
}^{\dagger }\left( \omega ^{\prime }\right) a_{2,\mathrm{out}}\left( \omega
\right) \right\rangle   \notag \\
&=&S_{1,e}+S_{1,o}+S_{2,e}+S_{2,o}+S_{\mathcal{G}}+S_{m}
\end{eqnarray}%
with
\begin{subequations}
\begin{eqnarray}
S_{1,e}&=&\frac{1}{2}\left[ \mathscr{T}_{5,1}\left( \omega \right) \mathscr{T}%
_{6,2}\left( -\omega \right) +\mathscr{T}_{6,1}\left( \omega \right)
\mathscr{T}_{5,2}\left( -\omega \right) \right] \text{,}~ S_{1,o}=\frac{1}{2}\left[\mathscr{T}_{5,3}\left( \omega \right) \mathscr{T}%
_{6,4}\left( -\omega \right) +\mathscr{T}_{6,3}\left( \omega \right)
\mathscr{T}_{5,4}\left( -\omega \right) \right] \text{,}  \\
S_{2,e}&=&\frac{1}{2}\left[ \mathscr{T}_{5,5}\left( \omega \right) \mathscr{T}%
_{6,6}\left( -\omega \right) +\mathscr{T}_{6,5}\left( \omega \right)
\mathscr{T}_{5,6}\left( -\omega \right) \right] \text{,}~  S_{2,o}=\frac{1}{2}\left[ \mathscr{T}_{5,7}\left( \omega \right) \mathscr{T}%
_{6,8}\left( -\omega \right) +\mathscr{T}_{6,7}\left( \omega \right)
\mathscr{T}_{5,8}\left( -\omega \right) \right] \text{,}   \\
S_{\mathcal{G}}&=&\frac{1}{2}\left[ \mathscr{T}_{5,10}\left( \omega \right) \mathscr{T}%
_{6,9}\left( -\omega \right) +\mathscr{T}_{6,10}\left( \omega \right)
\mathscr{T}_{5,9}\left( -\omega \right) \right] \text{,}~  S_{m}=\mathscr{T}_{5,12}\left( \omega \right) \mathscr{T}_{6,12}\left( -\omega
\right) \left( n_{m}+\frac{1}{2}\right) \text{,}
\end{eqnarray}
\end{subequations}
\end{widetext}
where $S_{j,e}$ and $S_{j,o}$ ($j=1,2$) represent the effects of the external and internal noises
to cavity $j$ rising from the optical vacuum fluctuations, respectively; $S_{%
\mathcal{G}}$ stands for the effect of the noise originating from the optical gain; $S_{m}$
represents the effect of the thermal noise to the mechanical modes.

Now we define noise-to-signal ratio ($\mathrm{NSR}$) in the first case as the ratio of
the integral of the output spectra $S_{2,\mathrm{out}}\left( \omega \right)$
and the output signal amplitude $\left\vert \alpha _{2,\mathrm{out}%
}\right\vert ^{2}$ to describe the quantity of the added noise in the output
port~\cite{otterstrom}. Experimentally, the noise under consideration will be detected by
a measurement device with a small bandwidth $2\Delta \omega $ around $\omega =0$, in this case $\mathrm{NSR}$ can be defined
by~\cite{mercier,otterstrom}

\begin{equation}
\mathrm{NSR}:=\frac{1}{\left\vert \alpha_{2,\mathrm{out}}\right\vert ^{2}}%
\int_{-\Delta \omega }^{\Delta \omega }d\omega S_{2,\mathrm{out}}\left(
\omega \right) \text{.}  \label{n1}
\end{equation}

Similarly, in the second case that the input filed is only injected
into cavity $2$, we can accordingly define $\widetilde{\mathrm{NSR}}$ to describe
the quantity of the added output noise as
\begin{equation}
\widetilde{\mathrm{NSR}}:=\frac{1}{\left\vert \tilde{\alpha}_{1,\mathrm{out}%
}\right\vert ^{2}}\int_{-\Delta \omega }^{\Delta \omega }d\omega \tilde{S}%
_{1,\mathrm{out}}\left( \omega \right) \text{,}  \label{n2}
\end{equation}%
where
\begin{widetext}
\begin{eqnarray}
\tilde{S}_{1,\mathrm{out}}\left( \omega \right)  &=&\frac{1}{2}\int \frac{%
d\omega ^{\prime }}{2\pi }\left\langle \tilde{a}_{1,\mathrm{out}}\left(
\omega \right) \tilde{a}_{1,\mathrm{out}}^{\dagger }\left( \omega ^{\prime
}\right) +\tilde{a}_{2,\mathrm{out}}^{\dagger }\left( \omega ^{\prime
}\right) \tilde{a}_{2,\mathrm{out}}\left( \omega \right) \right\rangle\notag \\
&=&\tilde{S}_{1,e}+\tilde{S}_{1,o}+\tilde{S}_{2,e}+\tilde{S}_{2,o}+\tilde{S}_{\mathcal{G}}+\tilde{S}_{m}
\end{eqnarray}
with
\begin{subequations}
\begin{eqnarray}
\tilde{S}_{1,e}&=&\frac{1}{2}\left( \tilde{\mathscr{T}}_{1,1}\left( \omega \right) \tilde{%
\mathscr{T}}_{2,2}\left( -\omega \right) +\tilde{\mathscr{T}}_{2,1}\left(
\omega \right) \tilde{\mathscr{T}}_{1,2}\left( -\omega \right) \right)\text{,}~
\tilde{S}_{1,o}=\frac{1}{2}\left( \tilde{\mathscr{T}}_{1,3}\left( \omega \right) \tilde{%
\mathscr{T}}_{2,4}\left( -\omega \right) +\tilde{\mathscr{T}}_{2,3}\left(
\omega \right) \tilde{\mathscr{T}}_{1,4}\left( -\omega \right) \right)
\\
\tilde{S}_{2,e}&=&\frac{1}{2}\left( \tilde{\mathscr{T}}_{1,5}\left( \omega \right) \tilde{%
\mathscr{T}}_{2,6}\left( -\omega \right) +\tilde{\mathscr{T}}_{2,5}\left(
\omega \right) \tilde{\mathscr{T}}_{1,6}\left( -\omega \right) \right)\text{,}~
\tilde{S}_{2,o}=\frac{1}{2}\left( \tilde{\mathscr{T}}_{1,7}\left( \omega \right) \tilde{%
\mathscr{T}}_{2,8}\left( -\omega \right) +\tilde{\mathscr{T}}_{2,7}\left(
\omega \right) \tilde{\mathscr{T}}_{1,8}\left( -\omega \right) \right)
 \\
\tilde{S}_{\mathcal{G}}&=&\frac{1}{2}\left( \tilde{\mathscr{T}}_{1,10}\left( \omega \right) \tilde{%
\mathscr{T}}_{2,9}\left( -\omega \right) +\tilde{\mathscr{T}}_{2,10}\left(
\omega \right) \tilde{\mathscr{T}}_{1,9}\left( -\omega \right) \right)\text{,}~
\tilde{S}_{m}=\tilde{\mathscr{T}}_{1,12}\left( \omega \right) \tilde{\mathscr{T}}%
_{2,12}\left( -\omega \right) \left( n_{m}+\frac{1}{2}\right) \text{.}
\end{eqnarray}%
\end{subequations}
\end{widetext}

\begin{figure}[htbp]
\centering
\includegraphics[width=8cm]{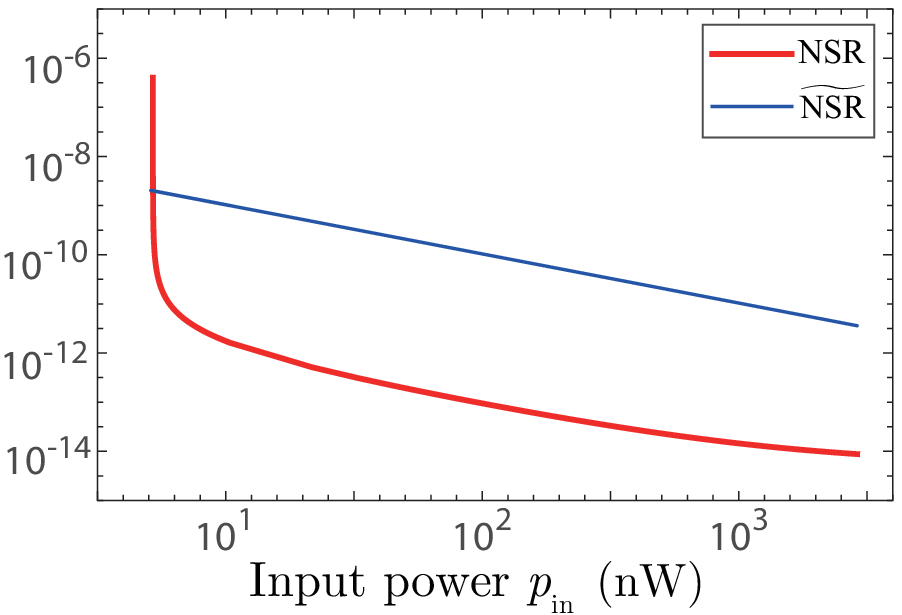} 
\caption{(Color online) $\mathrm{NSR}$ (red line) and $\widetilde{\mathrm{NSR}}$ (blue line) as a function of the
input power $p_{\mathrm{in}}$. We only plot $\mathrm{NSR}$ on the situation that
the system works on the upper branch of $T$ and the lower branch of $\widetilde{\mathrm{NSR}}$. The region of the input power $p_{\mathrm{in}}$ is the working region in Fig.~\protect\ref%
{Fig-2}(b), where $p_{\mathrm{in}%
}\in \lbrack 5.2\,\mathrm{nW},~2.83\,\mathrm{\mu W}]$. Here $n_{m}=100$, $\Delta \protect\omega/2\protect\pi=30\,\text{%
Hz,}$ and the other parameters are the same as that in Fig. 2(b).}
\label{Fig-n}
\end{figure}

For the reason that the analytical expressions of Eq.~(\ref{n1}) and Eq.~(%
\ref{n2}) are so complex, we numerically display NSR ($\widetilde{\mathrm{NSR}}$)
as a function of the input power $p_{\mathrm{in}}$ in Fig.~\ref{Fig-n}. In
the following simulations, we take the typical bandwidth $\Delta \omega/2\pi=30\,\text{Hz} $
as in the current experimental condition~\cite{mercier}. Moreover,
in order to clearly display the impact of the added noise on optical
directional amplification, we only focus on the added noise of the system
working on the upper branch of $T$ and the lower branch of $\tilde{T}$ with $p_{\mathrm{in}}\in \lbrack p_{%
\mathrm{in},l},~p_{\mathrm{in},u}] $ in Fig.~\ref{Fig-2}(b).

As shown in Fig.~\ref{Fig-n}, the maximum value of either $\mathrm{NSR}$ or $%
\widetilde{\mathrm{NSR}}$ is smaller than $10^{-6}$ in the working region in Fig.~%
\ref{Fig-2}(b). This means that the effects of the added noise in our proposal of optical nonreciprocal transmission with unidirectional amplification can be ignored.

\section{Conclusions}

In summary, it is found that assisted by the optical gain, the nonreciprocal
transmission with unidirectional amplification can be realized for a strong
optical input signal in our three-mode optomechanical system. The origin of
the optical amplification comes from the optical gain. An interesting
property of our system is that it simultaneously has high isolation ratio
and high transmission coefficient in a particular direction. Furthermore,
the expressions for the optimal transmission coefficient in the amplified
direction and the corresponding isolation ratio are analytically obtained.
However, there is a fact that should be stressed: the unidirectional
amplification in our system is sensitive to the power of input signal field,
and overcoming this issue is a new question and needs a future study.

\section{ACKNOWLEDGMENTS}

This work was supported by the Science Challenge Project (under Grant
No.~TZ2018003), the National Key R\&D Program of China under Grant
No.~2016YFA0301200, the National Natural Science Foundation of China (under
Grants No.~11774024, No.~11534002, No.~11874170, No.~11604096, No.~U1930402,
and No.~U1730449), and the Postdoctoral Science Foundation of China (under
Grant No. 2017M620593).

\end{document}